\documentclass[a4paper,fleqn,usenatbib]{emulateapj}
\usepackage{ae,aecompl}
\usepackage{graphicx}	
\usepackage{amsmath}	
\usepackage{amssymb,color}	

\newcommand{\Msun}{\ensuremath{\mathrm{M}_\odot}}

\newcommand{\R}{\textit{r}-process }

\slugcomment{draft v1}

\shorttitle{}
\shortauthors{Matsumoto, Ioka, Kisaka, and Nakar}

\begin{document}

\title{Is the Macronova in GW170817 Powered by the Central Engine?}

\author{Tatsuya Matsumoto\altaffilmark{1,2,3}}
\email{matsumoto@tap.scphys.kyoto-u.ac.jp}
\author{Kunihito Ioka\altaffilmark{4}}
\author{Shota Kisaka\altaffilmark{3,5}}
\author{Ehud Nakar\altaffilmark{6}}

\altaffiltext{1}{Department of Physics, Graduate School of Science, Kyoto University, Kyoto 606-8502, Japan}
\altaffiltext{2}{Racah Institute of Physics, Hebrew University, Jerusalem, 91904, Israel}
\altaffiltext{3}{JSPS Research Fellow}
\altaffiltext{4}{Center for Gravitational Physics, Yukawa Institute for Theoretical Physics, Kyoto University, Kyoto 606-8502, Japan}
\altaffiltext{5}{Department of Physics and Mathematics, Aoyama Gakuin University, Sagamihara, Kanagawa, 252-5258, Japan}
\altaffiltext{6}{The Raymond and Beverly Sackler School of Physics and Astronomy, Tel Aviv University, Tel Aviv 69978, Israel}

\begin{abstract}
The gravitational wave event GW170817 from a binary neutron star (NS) merger is accompanied by electromagnetic counterparts, and the optical and near-infrared emission is called a macronova (or kilonova). Although the radioactivity of synthesized \R elements is widely discussed as an energy source, its decisive evidence is not clearly shown yet. We discuss a macronova powered by the central engine activities such as jet activities and X-rays from the matter fallback, and show that the engine model allows much broader parameter spaces, in particular smaller ejecta mass ($\sim10^{-4}-0.01\,\Msun$) than the \R model. The blue and red macronovae are naturally explained by various combinations of the ejecta such as a cocoon and merger ejecta with the energy sources of jets and X-rays. The required energy injection is very similar to the X-ray excess observed in GRB 130603B with the power-law slope of $\sim-1.3$. The required lanthanoid fraction for the opacity can be also consistent with the Galactic one. Early or late multi-wavelength observations are crucial for revealing the central engine of short gamma-ray bursts and the \R nucleosynthesis.
\end{abstract}


\section{INTRODUCTION}\label{intro}
The LIGO-VIRGO collaboration detected the gravitational waves (GWs) from a binary neutron star (NS) merger for the first time \citep{Abbott+17}.
NS mergers have been expected to accompany electromagnetic signals with rich physical information.
Just after the GW detection, many follow-up observations in various wavelengths were performed \citep[][ and references therein]{ligo+17}.
The \textit{Fermi} and INTEGRAL satellites observed a gamma-ray signal $\sim1.7\rm{\,s}$ after the GW detection, which may reveal that some short gamma-ray bursts (sGRBs) originate from NS mergers \citep{ligo+17b,Goldstein+17,Savchenko+17}.
In other wavelengths such as X-ray, ultra-violet, optical, near infrared (NIR), and radio bands, electromagnetic counterparts were subsequently detected.
These detections associated with the GW event open up the era of multi-messenger astronomy.

In this work, we focus mainly on the counterparts in optical and NIR bands \citep{Arcavi+2017,Chornock+17,Coulter+17,Cowperthwaite+17,Diaz+17,Drout+2017,Evans+2017,Kasliwal+2017,Kilpatrick+17,McCully+17,Nicholl+17,Pian+2017,Shappee+17,Smartt+2017,Soares-Santos+17,Tanvir+17,Tominaga+2017,Utsumi+17,Valenti+17}.
The optical and NIR counterparts of GW170817, SSS17a (or AT2017fgo), were first discovered by the Swope Supernovae Survey about half a day after the merger \citep{Coulter+17}.
SSS17a showed a bright optical emission at first few days, and its spectra are well fitted by a blackbody.
The bolometric luminosity and the temperature are evaluated as $L\simeq7\times10^{41}\,\rm{erg\,s^{-1}}$ and $T\simeq7000\,\rm{K}$ at 1 day.
Afterwards, the optical emission decayed and transitioned to an NIR emission in a week.
Although the spectrum began to deviate from a blackbody, the luminosity and temperature are evaluated as $L\simeq8\times10^{40}\,\rm{erg\,s^{-1}}$ and $T\simeq3000\,\rm{K}$ at 1 week.

Theoretically, an electromagnetic counterpart following a binary NS merger was predicted a few decades ago, and it was named as a macronova \citep{1998ApJ...507L..59L,Kulkarni2005}, or kilonova \citep{Metzger+2010}.
Binary NS mergers are one of the most promising production sites of the heavy elements \citep{Lattimer&Schramm1974,Symbalisty&Schramm1982,Eichler+1989}.
The coalescence of NSs ejects neutron-rich matter, where heavy elements are synthesized by the rapid neutron capture, so-called the \textit{r}-process.
The produced \R elements are unstable and radioactively decay to heat up the ejecta, which results in optical and NIR emissions.
In this work, we call this scenario as the \textit{r-process model}.

In order to reproduce the SSS17a's early-optical and late-NIR emissions in the \R model, at least two emission regions with different properties are required in the polar or radial direction \citep{Smartt+2017,Waxman+2017}.
These optical and NIR emissions are called blue and red macronovae, respectively.
In the \R model, since the luminosity is proportional to the ejecta mass, we can estimate the ejecta mass from the observed bolometric luminosity.
The detailed light curve or spectral modeling also gives the ejecta mass as $M_{\rm{ej}}^{\rm{blue}}\simeq0.02\,\Msun$ \citep{Arcavi+2017,Chornock+17,Cowperthwaite+17,Drout+2017,Kasen+17,Kilpatrick+17,McCully+17,Nicholl+17,Smartt+2017} and $M_{\rm{ej}}^{\rm{red}}\simeq0.03\,\Msun$ \citep{Chornock+17,Cowperthwaite+17,Drout+2017,Kasen+17,Kilpatrick+17,Tanaka+17,Utsumi+17} for the blue and red macronovae, respectively.
Given the ejecta masses, the different timescales of the blue ($\sim{\,\rm{day}}$) and red ($\sim{7\,\rm{days}}$) macronovae suggest different opacities of  $\kappa\sim0.1-1{\,\rm{cm^{2}}\,g^{-1}}$ and $\kappa\sim1-10{\,\rm{cm^{2}}\,g^{-1}}$, respectively.
The difference of the opacities reflects the different abundances of the synthesized \R elements \citep[in particular, lanthanoids in this model,][]{Kasen+2013,Tanaka&Hotokezaka2013,Tanaka+2017}.
Numerical simulations of binary NS mergers and nucleosynthesis also suggest some distribution of the ejecta properties \citep{Metzger&Fernandez2014,2014MNRAS.443.3134P,Lippuner+2017,Shibata+2017}.

Although the \textit{r}-process model is broadly accepted as the standard model, it may contain some uncomfortable tensions;
\begin{enumerate}
\item First, the total ejecta mass required by the \textit{r}-process model, $M_{\rm{ej}}={M_{\rm{ej}}^{\rm{blue}}+M_{\rm{ej}}^{\rm{red}}\simeq0.05\,\Msun}$, seems to be relatively larger than that expected by some numerical calculations.
Currently, the binary merger calculations show the ejecta mass at the onset of a merger (the dynamical ejecta) $M_{\rm{ej}}\lesssim0.01-0.02\,\Msun$ \citep{Hotokezaka+2013,Sekiguchi+2015,Sekiguchi+2016}, except for ones with extremely large mass ratios of $q<0.7$ \citep{Dietrich+2017}, which are not observed in the Galactic binary pulsars \citep{Tauris+2017}.
The post-merger ejecta such as the viscously-driven outflow may contain rather larger mass of $\gtrsim0.01\,\Msun$ \citep{Fujibayashi+2017b,Shibata+2017}, but its velocity ($v\simeq0.05-0.15\,c$, where $c$ is the speed of light) does not seem so large as that suggested by the observations ($v\simeq0.3\,c$, in particular, for the blue macronova).
We also remark that the exact values may depend on the prescription of the neutrino transfer and the viscosity.
Intriguingly, massive ejecta were also suggested by the macronova candidates accompanying GRBs 050709, 060614, and 130603B \citep[$\sim0.03-0.1\,\Msun$,][]{Berger+2013,Tanvir+2013,Yang+2015,Jin+2016}.

\item Given the merger rate estimated from this event, the necessary ejecta mass in the \textit{r}-process model could result in a larger production rate of the \textit{r}-process elements than that suggested by the solar abundance \citep[see also][]{Cowperthwaite+17}.
Combining the observed NS merger rate of ${\cal{R}}\simeq1540_{-1220}^{+3220}{\,\rm{Gpc^{-3}\,yr^{-1}}}$ \citep{Abbott+17} with the estimated ejecta mass of $M_{\rm{ej}}\simeq0.05\,\Msun$ gives the  \textit{r}-process production rate in a galaxy of $\dot{M}_{r}={\cal{R}}M_{\rm{ej}}/n_{\rm{Gal}}\simeq7.7_{-6.1}^{+16.1}\times10^{-6}\,\Msun\,\rm{yr^{-1}}$, where we use the number density of galaxies of $n_{\rm{Gal}}\simeq0.01\,\rm{Mpc^{-3}}$.
The central value of the above production rate is ten times larger than the Galactic production rate of $\simeq7\times10^{-7}\,\Msun{\rm{\,yr^{-1}}}$ for the \R elements with mass numbers $A\gtrsim100$ \citep{2000ApJ...534L..67Q}.
Of course, we should remark again that the adopted merger rate is determined only by GW170817.

\item It is also uncertain whether the ejecta required to explain SSS17a can reproduce the Galactic \R abundance pattern or not.
In order to explain the solar abundance pattern, the lanthanoid mass fraction of $X_{\rm{Lan}}\sim10^{-1.5}$ is required for the ejecta composed of \R elements.
However, for instance, \cite{Waxman+2017} conclude that the observation of SSS17a suggests the ejecta's lanthanoid mass fraction of $X_{\rm{Lan}}\sim10^{-3}$.

\item Compared with the other sGRBs, the light curve of SSS17a suggests that there may be a diversity in the macronova luminosity \citep{Fong+2017,Gompertz+2017}.
The luminosity dispersion could have a range of two-orders of magnitudes.
In the \R model, if the previous macronovae were also produced by binary NSs and had similar spectra to this event, the dispersion of the luminosity reflects the diversity of the ejecta mass.
However, such a broad diversity in ejecta mass may not be produced in particular by the dynamical coalescence phase, which depends on the binary's total mass and mass ratio showing narrow distributions among the Galactic binary pulsars \citep{Tauris+2017}.
The post-merger ejecta such as the viscously-driven outflow may explain the diversity due to the differences of the magnetic fields or the collapse time of remnant hypermassive NSs to black holes (BHs), but the numerical calculations are not conclusive yet.

\item The gamma-, X-ray, and radio observations suggest the existence of energetic ejecta such as a relativistic jet or a mildly-relativistic cocoon \citep{ligo+17b,Alexander+17,Evans+2017,Fong+2017,Goldstein+17,Haggard+17,Hallinan+17,Ioka&Nakamura2017,Kasliwal+2017,Margutti+17,Murguia-Berthier+17,Savchenko+17,Troja+2017}.
These ejecta can easily inject energy into the macronova components and affect the emissions.

\end{enumerate}

Then, it is worth considering another energy source rather than the radioactive heating by the \R elements, and discussing whether the energy source can explain the observation or not.
In this work, we consider the energy injection from the central engine activities as the energy source, and call this model as the \textit{engine model}.
Different from the \R model, the energy source is decoupled with the ejecta mass, so that wider ranges of the ejecta mass and opacity are allowed in the engine model. 
In particular, the engine model can reproduce the macronova by less ejecta mass than that needed by the \textit{r}-process model, and may resolve the above concerns 1 and 2.
Furthermore, since the jet activities actually show the diversity in the luminosity or energy, the diversity of macronovae (concern 4) can be also explained by the difference of the injection energy.

So far, the energy injection mechanisms have been discussed and applied to the macronova in GRB 130603B \citep{Kisaka+2015,Kisaka+2016}.
Some sGRBs show long-lasting jet activities such as the extended emissions \citep[$\sim10^2\,\rm{s}$,][]{Barthelmy+2005} and the plateau emissions \citep[$\sim10^{4-5}\,\rm{s}$,][]{Gompertz+2013,Gompertz+2014}, which may result from the activities of the central engine \citep{Ioka+2005}, such as BHs \citep{Kisaka&Ioka2015} or NSs \citep{2008MNRAS.385.1455M}.
If a rapidly rotating magnetar is formed as the central engine, its large spin-down energy can produce a bright (or even brighter) optical emission like a macronova \citep{Yu+2013,Metzger&Piro2014}.
Since a long-lived magnetar makes the emission too bright, the magnetar likely collapses to a BH in a short timescale.
After the collapse, the BH launches a jet and injects energy into the ejecta by radiations or shocks, reproducing the macronova \citep{Kisaka+2015}.
In addition, a mysterious X-ray excess with comparable duration ($\simeq7\,\rm{days}$) and luminosity ($\simeq10^{41-42}\,\rm{erg\,s^{-1}}$) was also detected in GRB 130603B \citep{Fong+2014}.
If the X-ray excess was emitted quasi-isotropically and absorbed by the ejecta, the reprocessed emission of the X-rays is also able to produce the macronova emission \citep{Kisaka+2016}.
For SSS17a, \cite{Ioka&Nakamura2017} show that the blue macronova could be powered by the prompt jet only \citep[see also,][]{Piro&Kollmeier2017}.
\cite{Yu&Dai2017} also discuss the energy injection from the spin-down of a remnant hypermassive NS, although the NS should keep unrealistically small magnetic fields $B\sim10^{11-12}\,\rm{G}$.

The observations of SSS17a supply very rich data in contrast to GRB 130603B with only one-epoch detection and other macronova candidates.
In this work, we extend the engine model and test whether the energy injection can explain the extensively observed macronova, SSS17a or not.
The structure of this paper is as follows.
In section \ref{engine model}, we discuss the basic concept of our engine model and show that this model can reproduce the observed macronova with order-of-magnitude estimations.
The engine model allows a large parameter space to explain this event.
We argue that some parameter sets are actually motivated from numerical simulations.
Then, in section \ref{LIGHT CURVE MODEL}, we construct simple light curve models and show that the models agree with the observation, with smaller ejecta mass than that required in the \R model, and with the observationally-motivated energy injection channel. 
Finally we discuss the implications of our study and the prospect of the future observations in section \ref{DISCUSSION}.

\section{ENGINE MODEL}\label{engine model}
The observed macronova showed a transition from the optical ($\sim1\,\rm{day}$) to the NIR emissions ($\sim1\,\rm{week}$).
We call the early-optical and late-NIR emissions as blue and red macronovae, respectively.
We estimate the emission radii of the blue and red macronovae by using the observables \citep{Drout+2017,Kasliwal+2017,Kilpatrick+17}.
Let us consider an ejecta emitting photons and subtending a fraction $\Omega$ of the solid angle. 
By assuming blackbody radiation, the emission radius with the luminosity $L$ and temperature $T$ is given by 
\begin{eqnarray}
R=\sqrt{\frac{L}{4\pi\Omega\sigma{T^4}}},
\end{eqnarray}
where $\sigma$ is the Stefan-Boltzmann constant.
With the observables, the emission radii of the blue and red macronovae are estimated by
\begin{eqnarray}
R_{\rm{blue}}&\simeq&9.0\times10^{14}{\rm{\,cm\,}}\biggl(\frac{T}{7000\,\rm{K}}\biggl)^{-2}\nonumber\\
&&\biggl(\frac{L}{7\times10^{41}{\rm{\,erg\,s^{-1}}}}\biggl)^{1/2}\Omega_{0.5}^{-1/2},\\
R_{\rm{red}}&\simeq&1.7\times10^{15}{\rm{\,cm\,}}\biggl(\frac{T}{3000\,\rm{K}}\biggl)^{-2}\nonumber\\
&&\biggl(\frac{L}{8\times10^{40}{\rm{\,erg\,s^{-1}}}}\biggl)^{1/2}\Omega_{0.5}^{-1/2},
\end{eqnarray}
where we use $\Omega=0.5\,\Omega_{0.5}$.
By dividing each radius by each timescale, we infer the expanding velocities of $v_{\rm{blue}}\simeq0.3\,c$ and $v_{\rm{red}}\simeq0.1\,c$, respectively, where $c$ is the speed of light.
Note that after 7 days, the red macronova is no longer approximated by a blackbody spectrum \citep{Chornock+17,Kilpatrick+17,Nicholl+17,Pian+2017,Shappee+17}.  
The difference of the photospheric velocities suggests that the blue and red macronovae are powered by the different emission regions along the polar or radial direction.

The large range of the emission timescale from 1 day to 7 days suggests that there are at least two emission regions contained by more than one component of ejecta.
It should be noted that the two emission regions are not necessarily contained by two ejecta components separately.
For example, the observed light curve of the macronova is well fitted by a single-component-ejecta models \citep[see][]{Smartt+2017,Waxman+2017}.
Before discussing our model, we define the meaning of the word ``component" of ejecta.
In the following, we discuss the properties of various ejecta.
When we can specify an ejecta based on its physical origin such as the dynamical ejecta or post-merger ejecta (see below), we call each ejecta as ``component".
These components have some continuous opacity, density, and velocity distributions, which produce emission regions with different properties such as emission timescales. 

\begin{figure}
\begin{center}
\includegraphics[width=60mm, angle=270]{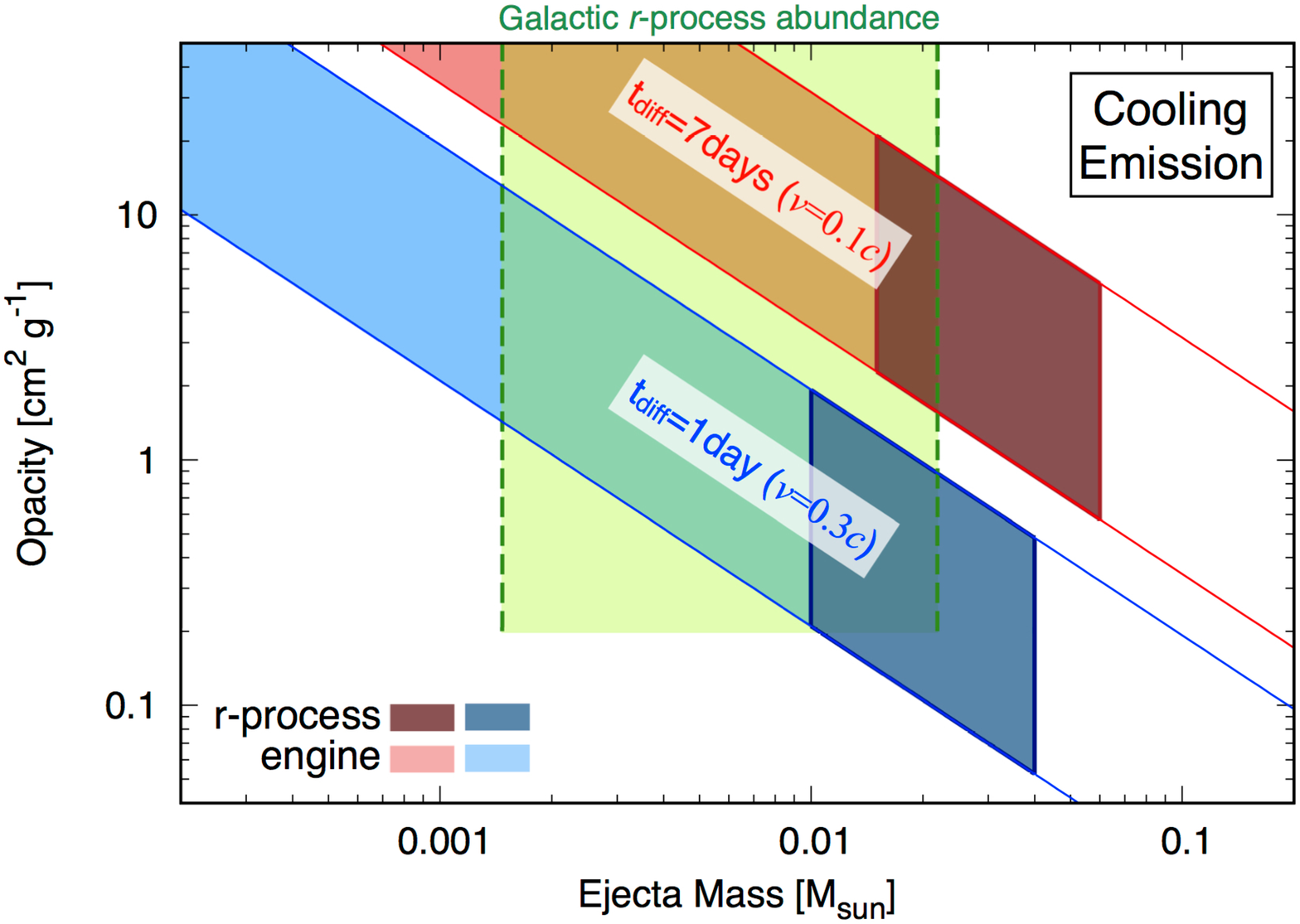}
\caption{The required ejecta mass and opacity regions to reproduce the observed blue ($t_{\rm{diff}}\simeq1\,\rm{day}$) and red ($t_{\rm{diff}}\simeq7\,\rm{days}$) macronovae with cooling emissions in the engine model. The dark-red and dark-blue regions denote the mass and opacity suggested in the \R model. The green shaded region shows the required total ejecta mass to explain the Galactic \R abundance with the event rate estimated by GW170817. The engine model allows much larger parameter ranges than the \R model.}
\label{fig km}
\end{center}
\end{figure}

In the engine model, emissions are not powered by the radioactive decay of \R elements, but by the activity of the central engine, which allows large parameter spaces of the ejecta mass and opacity.
In order to see this, we first consider that the injected energy is released by the photon diffusion, so-called the cooling emission.
The characteristic timescale of the cooling emission is set by the condition of $\tau=c/v$, where $\tau$ and $v$ are the optical depth and velocity of the ejecta, respectively.\footnote{The diffusion timescale also gives the peak timescale of the emission in the \R model.}
The diffusion time of the ejecta with an opacity $\kappa$ and mass $M$ is given by \citep{Arnett1980}
\begin{eqnarray}
t_{\rm{diff}}\simeq\sqrt{\frac{\xi\kappa{M}}{\Omega{vc}}},
\label{t_diff 1}
\end{eqnarray}
where $\xi$ is a numerical factor reflecting the density structure of the ejecta.
For instance, \cite{Arnett1980} originally evaluated the factor as $\xi\simeq2/13.7$, and we use $\xi=3/4\pi\simeq0.24$ in the one-zone model and $\xi\simeq0.026$ for ejecta with a power-law density distribution (see section \ref{LIGHT CURVE MODEL}).
The observed macronova showed a large timescale range of $1-7\,\rm{days}$ which may be difficult to reproduce by a single combination of the opacity and mass.
Then, we consider at least two components of ejecta with different values of the product $\kappa{M}$ or a single component with a polar or radial distribution of $\kappa{M}$.
By using the emission timescales of 1 and 7 days, and the ejecta velocities of $v_{\rm{blue}}=0.3\,c$ and $v_{\rm{red}}=0.1\,c$, the products are constrained as 
\begin{eqnarray}
\kappa{M}&\simeq&2.1\times10^{-3}{\,\rm{cm^2\,g^{-1}}\,\Msun}\nonumber\\
&&\biggl(\frac{v_{\rm{blue}}}{0.3\,c}\biggl)\biggl(\frac{t_{\rm{diff}}}{1\,{\rm{day}}}\biggl)^2\biggl(\frac{\xi}{3/4\pi}\biggl)^{-1}\Omega_{0.5},
   \label{kappaM cool blue}\\
\kappa{M}&\simeq&3.4\times10^{-2}{\,\rm{cm^2\,g^{-1}}\,\Msun}\nonumber\\
&&\biggl(\frac{v_{\rm{red}}}{0.1\,c}\biggl)\biggl(\frac{t_{\rm{diff}}}{7\,{\rm{day}}}\biggl)^2\biggl(\frac{\xi}{3/4\pi}\biggl)^{-1}\Omega_{0.5},
\end{eqnarray}
for the blue and red macronovae, respectively.
In Fig. \ref{fig km}, we show the constraints on the ejecta mass and opacity imposed by the observed timescales.
The red and blue shaded regions show the above constraints on $\kappa{M}$ taking the uncertainty of the coefficient $\xi$ in Eq. \eqref{t_diff 1} into account.
The dark-red and dark-blue regions show the parameter spaces in the \R model.
In order to explain the observations in the engine model, the mass should be smaller than that required in the \R model.
The green shaded region denotes the total ejecta mass required to explain the Galactic \R abundance with mass numbers $A\gtrsim100$ by only binary NS mergers with the event rate estimated by this event, ${\cal{R}}\simeq1540_{-1220}^{+3220}{\,\rm{Gpc^{-3}\,yr^{-1}}}$ \citep{Abbott+17}.
Since the low opacity value of $\sim0.2\,\rm{cm^2\,g^{-1}}$ is realized by hydrogen or iron without \R elements, the green shaded region has a boundary at the low opacity. 
In principle, the ejecta in the blue and red shaded regions can reproduce the observed blue and red macronovae, respectively, in the engine model.

\subsection{Ejecta}\label{Ejecta}
In the following, we discuss possible ejecta components, which are motivated by the recent numerical simulations, although the engine model allows more general configurations as in Fig. \ref{fig km}.

\if{
\begin{table}
\begin{center}
\caption{The possible ejecta of binary mergers}
\label{table ejecta}
\begin{tabular}{crr}
\tableline\tableline
Ejecta&$\kappa{M}$[$\rm{\,cm^2\,g^{-1}}\,\Msun$]&$v/c$\\
\tableline
Merger ejecta&0.1&0.1\\
Choked jet cocoon&0.0025&0.3\\
Successful jet cocoon&0.0025&0.3\\
\tableline
\end{tabular}
\end{center}
\end{table}
}\fi

\subsubsection{Merger Ejecta}\label{Merger Ejecta}
One is the \textit{merger ejecta}. 
The merger ejecta includes the dynamical ejecta \citep{Hotokezaka+2013,Sekiguchi+2015,Sekiguchi+2016} and the post-merger ejecta such as the neutrino-driven winds \citep{2009ApJ...690.1681D,Wanajo+2012,2014MNRAS.443.3134P,2017ApJ...846..114F} and more importantly viscously-driven outflows \citep{2013MNRAS.435..502F,2015MNRAS.446..750F,2015MNRAS.449..390F,2014PhRvD..90d1502K,2015PhRvD..92f4034K,2015ApJ...809...39G,2017PhRvD..95f3016C,2017PhRvD..95h3005S,Siegel&Metzger2017,Fujibayashi+2017b}.
For example, we can consider the following set of parameters: $\kappa_{\rm{e}}=10\,\rm{cm^{2}\,g^{-1}}$, $M_{\rm{e}}=10^{-2}\,\Msun$, and $v_{\rm{e}}=0.1\,c$.
It should be noted that as long as the product of $\kappa_{\rm{e}}$ and $M_{\rm{e}}$ has the same value, the resulting emission has a similar signature in the engine model.

The dynamical ejecta is produced by the shock heating and the tidal interaction at the onset of the merger.
The ejecta mass and velocity depend on the equation of state of nuclear matter, the binary mass ratio, and so on.
Numerical relativity simulations show typical values of $M_{\rm{e}}\sim10^{-3}-10^{-2}\,\Msun$ and $v_{\rm{e}}\simeq0.1-0.2\,c$.\footnote{We do not consider the ultra-relativistic ejecta from the shock breakout of NSs \citep{Kyutoku+2014}.}
The dynamical ejecta may have different opacities depending on the angluar direction \citep{Wanajo+2014,Tanaka+2017}.
In the equatorial direction, the ejecta is launched mainly by the tidal force and has a low electron fraction of $Y_e\sim0.1-0.2$ due to the inefficient heating.
In such a very neutron rich ejecta, the nucleosynthesis advances up to the third peak and produces lanthanoid elements, which results in the opacity as high as $\kappa_{\rm{e}}\sim10\,\rm{cm^{2}\,g^{-1}}$.
On the other hand, shock heating also drives ejecta quasi-spherically and raises its electron fraction $Y_e\gtrsim0.25$.
Then, the shock-heated dynamical ejecta has rather small opacity of $\kappa_{\rm{e}}\sim0.1-1\,\rm{cm^{2}\,g^{-1}}$ \citep{Sekiguchi+2015,Sekiguchi+2016,Wanajo+2014,Tanaka+2017}.
In this case, the opacity smoothly increases from the polar to the equatorial direction.

The post-merger ejecta is launched by the merger remnant object and the accretion disk into the polar direction.
The ejecta is less neutron rich and has a small opacity $\kappa_{\rm{e}}\sim0.1-1\,\rm{cm^{2}\,g^{-1}}$.
The typical velocity and mass are $v_{\rm{e}}\simeq0.05\,c-0.15\,c$ and $M_{\rm{e}}\sim10^{-2}\,\Msun$ \citep{Siegel&Metzger2017,2017ApJ...846..114F,Fujibayashi+2017b,Shibata+2017}, but these values depend on the neutrino transport and the viscosity prescription in numerical simulations.
The post-merger ejecta is located in the inner region of the dynamical ejecta because its velocity is smaller than that of the dynamical ejecta.

\subsubsection{Cocoons}\label{Cocoons}
The other is the \textit{cocoon}.
When a relativistic jet is launched from the central engine and propagates in the merger ejecta on the polar axis, the jet energy is dissipated and injected into the cocoon.
The duration of the engine activity determines whether the prompt jet successfully breaks out of the merger ejecta or not.
Now successful and choked jet scenarios are proposed in order to explain the prompt gamma-ray emission and the radio and X-ray afterglows associated with GW170817 \citep{Kasliwal+2017,Gottlieb+2017b,Ioka&Nakamura2017,Lazzati+2017,Mooley+2017,Margutti+2018,Nakar&Piran2018}.
In this event, both models predict the cocoon with similar properties such as the mass and velocity.
When the jet drills through the merger ejecta successfully, the cocoon also breaks out of the ejecta and expands isotropically.\footnote{We mainly focus on the dominant, sub-relativistic cocoon, not on the mildly-relativistic cocoon \citep{2017ApJ...834...28N,Gottlieb+2017}.}
Even when the jet is choked within the merger ejecta, if a sufficient energy is injected into the cocoon, it breaks out of the ejecta \citep{Gottlieb+2017b}.

The cocoon parameters are evaluated as follows.
The dissipated jet energy is evaluated by $E_{\rm{jet}}\sim{}L_{\rm{j}}t_{\rm{j}}=5\times10^{50}\,{\rm{erg}}\,(L_{\rm{j}}/5\times10^{50}{\,\rm{erg\,s^{-1}}})(t_{\rm{j}}/1{\,\rm{s}})$, where $L_{\rm{j}}$ and $t_{\rm{j}}$ is the geometrically-corrected jet luminosity and the duration of the prompt-jet launching time, respectively.
The cocoon mass is determined by the volume swept by the cocoon's shock.
In the case of short GRB jets, differently from long GRB jets, the cocoon mass is evaluated by $M_{\rm{c}}\sim5\times10^{-3}\,\Msun\,({M_{\rm{e}}}/0.01\,\Msun)(\theta_{\rm{j}}/0.3)$, where $M_{\rm{e}}$ and $\theta_{\rm{j}}$ are the merger ejecta's mass and the jet opening angle, respectively \citep{Ioka&Nakamura2017}.
The adopted jet opening angle $\theta_{\rm{j}}=0.3{\,\rm{rad}}\simeq17^\circ$ is consistent with the mean angle suggested by the observations \citep[$\theta_{\rm{j}}\simeq16^\circ\pm10^\circ$,][]{Fong+2015}, which could also explain the observations of sGRB 170817A, blue macronova, X-ray and radio-afterglows following GW170817 for the successful jet breakout \citep{Ioka&Nakamura2017}.
Then, the velocity is given by $v_{\rm{c}}\sim\sqrt{2E_{\rm{jet}}/M_{\rm{c}}}\sim0.3\,c\,(E_{\rm{jet}}/5\times10^{50}{\,\rm{erg}})^{1/2}(M_{\rm{c}}/5\times10^{-3}\,\Msun)^{-1/2}$.
The cocoon has likely rather small opacity of $\kappa_{\rm{c}}\simeq0.5\,\rm{cm^{2}\,g^{-1}}$, because it is made of the polar directed merger ejecta, which is dominated by high electron fraction ejecta of $Y_e>0.25$ and synthesizes less opaque elements \citep[][and see also section \ref{Merger Ejecta}]{Sekiguchi+2015,Sekiguchi+2016,Wanajo+2014,Tanaka+2017}.
It should be noted again that only the product $\kappa_{\rm{c}}M_{\rm{c}}$ is important to reproduce the light curve in the engine model.

\subsection{Energy Sources}\label{Energy Sources}
We consider an alternative to \textit{r}-process radioactivity for the energy source of the blue and red macronovae: the central engine activities.
We discuss two kinds of observationally-motivated energy sources from the engine activities and evaluate the emission timescales of the ejecta.

\subsubsection{Jets}
A jet launched by the central engine can inject energy into ejecta in the polar direction by high energy radiations or shocks.
Since a cocoon is formed near the polar axis, we consider that the cocoon is heated up by the energy injection from the jet activities and powers a cooling emission.

In addition to the prompt emissions, some sGRBs show extended and plateau emissions \citep{Kisaka+2015,Kisaka+2017,Ioka&Nakamura2017}, which may result from the long-lasting jet activities.
These emissions are too dim to be detected by an off-axis observer in this event.
Recently, \cite{Kisaka+2017} investigated 65 sGRBs' X-ray light curves in \textit{Swift}/BAT and XRT data, and found that the typical (geometrically-corrected)\footnote{\cite{Kisaka+2017} studies the isotropic radiated energy $E_{\rm{iso,rad}}$, and the injection energy is evaluated by $E_{\rm{in}}{\sim}E_{\rm{jet}}\sim(\theta_{\rm{j}}/0.3)^2(\eta/0.1)^{-1}E_{\rm{iso,rad}}$, where $\eta$ is the emission efficiency.} injection energy and time are $E_{\rm{in}}(\sim{E_{\rm{jet}}})\sim10^{48-51}\,\rm{erg}$ and $t_{\rm{in}}\sim10^2{\,\rm{s}}$ for the extended emissions, and $E_{\rm{in}}\sim10^{47-51}{\,\rm{erg}}$ and $t_{\rm{in}}\sim10^{4-5}{\,\rm{s}}$ for the plateau emissions, respectively.
Since the injection energy is less than that injected by the prompt jet, the cocoon is not accelerated any more.
It should be noted that since a successful prompt jet punches a hole in the ejecta, the long-lasting jets should have a larger opening angle than that of the prompt jet in order to interact with the cocoon.

Due to the heating, the jet-powered cocoon radiates photons for a day.
The characteristic photon diffusion timescale is given by Eq. \eqref{t_diff 1},
\begin{eqnarray}
t_{\rm{diff}}\simeq1.1{\rm{\,day\,}}\kappa_{\rm{c,0.5}}^{1/2}M_{{\rm{c}},0.005}^{1/2}v_{{\rm{c}},0.3}^{-1/2}\Omega_{0.5}^{-1/2}\biggl(\frac{\xi}{3/4\pi}\biggl)^{1/2},
\label{t_diff cocoon}
\end{eqnarray}
where we use $\kappa_{\rm{c}}=0.5\,\kappa_{\rm{c,0.5}}\rm{\,cm^2\,g^{-1}}$, $M_{\rm{c}}=5\times10^{-3}\,M_{\rm{c,0.005}}\,\Msun$, and $v_{\rm{c}}=0.3\,c\,v_{\rm{c},0.3}$.
We define the fraction of the solid angle subtended by the jet-powered ejecta as $\Omega=1-\cos\theta$, where $\theta$ is the half apex angle of the ejecta (see Fig. \ref{fig picture}).
We consider the cocoon in a one-zone model (see below Eq. \eqref{t_diff 1} for the fiducial values of $\xi$).
The luminosity at $t=t_{\rm{diff}}$ is estimated by dividing the internal energy by the diffusion time, $L\simeq{E_{\rm{int}}(t=t_{\rm{diff}})}/t_{\rm{diff}}$.
Until the photons start to diffuse out, the internal energy suffers from the adiabatic cooling and decreases as $E_{\rm{int}}(t)=E_{\rm{in}}(t/t_{\rm{in}})^{-1}$.
Then, the diffusion luminosity is given by $L\sim{E_{\rm{in}}t_{\rm{in}}}/t_{\rm{diff}}^2$.
Therefore, by the energy injection of $E_{\rm{in}}t_{\rm{in}}\sim7\times10^{51}\,\rm{erg\,s}$, the diffusion luminosity reaches $L\sim10^{42}\,\rm{erg\,s^{-1}}$ at $t_{\rm{diff}}\sim1\,\rm{day}$ in optical bands.
Finally, we remark that a similar energy injection scenario is discussed in \cite{Kasliwal+2017} (see their supplementary material), where the engine-driven wind injects energy into the ejecta.

\subsubsection{Long-Lasting X-ray Luminosity}\label{Long-Lasting X-ray Luminosity}
We may have another energy source, which is motivated by the X-ray observation of GRB 130603B.
This event showed a mysterious X-ray excess with a long duration of $\sim7\,\rm{days}$, and a luminosity larger than the extrapolation from its afterglow and showing a power-law temporal decaying of $L_{\rm{X}}(t)\propto{t}^{-\alpha_{\rm{X}}}$\citep{Fong+2014}, where $\alpha_{\rm{X}}$ is the temporal index. 
Such an X-ray excess could be produced by the fallback accretion onto the central engine \citep{2007MNRAS.376L..48R,2009MNRAS.392.1451R,Kisaka&Ioka2015}. 
More interestingly, the excess X-ray luminosity is comparable with the luminosity of the NIR macronova associated with GRB 130603B.
If the X-ray excess emission is quasi-isotropic and absorbed by the merger ejecta, the reprocessed NIR photons can reproduce the NIR macronova without introducing any other energy sources such as the radioactive energy \citep{Kisaka+2016}.

We consider that the cocoon and merger ejecta are irradiated by the long-lasting X-ray emission, which is similar to the mysterious X-ray excess observed in GRB 130603B.
Then, the ejecta produce optical and NIR emissions for $\sim1-7\,\rm{days}$ by reprocessing X-rays, which are observed as the macronova.
Note that when jet activities inject less energy than the X-ray excess, the reprocessed emission dominates the cocoon emission.
Even without the cocoon, the irradiated merger ejecta may also produce the blue macronova of $\sim1\,\rm{day}$, if some part of the ejecta has small $\kappa{M}$ (see below).
For instance, the opacity of the merger ejecta may distribute from a large value ($\kappa\sim10\,\rm{cm^2\,g^{-1}}$) at the equatorial plane to a low value ($\kappa\sim0.1-1\,\rm{cm^2\,g^{-1}}$) at the polar axis (see also section \ref{Merger Ejecta}).
A distribution of $\kappa{M}$ is also realized even with a constant opacity for the whole ejecta, when the ejecta mass is distributed along the polar or radial direction.
In principle, the central engine activity can last for a long time of $\sim7\rm{\,days}$ and isotropically irradiate the ejecta by X-rays with the luminosity of $L_{\rm{X}}\sim10^{41-42}\,\rm{erg\,s^{-1}}$ \citep{Fong+2014,2016ApJ...833..151F,Kisaka+2017}.
Because the bound-free opacity of X-rays is much larger than that of optical and NIR photons, the ejecta absorb the X-rays and reprocess them into lower energy photons.
We do not observe the X-ray excess in this event, because we see GW170817 from off-axis angle as suggested by the prompt emission and afterglow observations, and the X-rays are absorbed by the merger ejecta.

For the reprocessed photons to escape from the ejecta, the ejecta should be ``diffusively thin".
Moreover, for the diffusing photons to show a blackbody spectrum, the ejecta should also be ``optically thick".
It should be noted that these two conditions are defined by different two concepts.
When the diffusion time of the ejecta is shorter than the dynamical timescale, photons can diffuse out of the ejecta after multiple scatterings.
This situation is called as diffusively thin, and expressed as
\begin{eqnarray}
t>t_{\rm{diff}}\simeq4.0{\rm{\,day\,}}\kappa_{\rm{e,10}}^{1/2}M_{\rm{e,0.01}}^{1/2}v_{{\rm{e,0.1}}}^{-1/2}\Omega_{0.5}^{-1/2}\biggl(\frac{\xi}{0.026}\biggl)^{1/2},
\label{t_diff m}
\end{eqnarray}
where we normalize $\xi$ with 0.026 by taking the density structure into account (see section \ref{LIGHT CURVE MODEL}).
We use $\kappa_{\rm{e}}=10\,\kappa_{\rm{e,10}}\rm{\,cm^2\,g^{-1}}$, $M_{\rm{e}}=10^{-2}\,M_{\rm{e,0.01}}\,\Msun$, and $v_{\rm{e}}=0.1\,c\,v_{{\rm{e}},0.1}$.
The diffusion time of the cocoon is given in Eq. \eqref{t_diff cocoon}.
In the optically thick ejecta, the photons are scattered and thermalized.
The timescale when the ejecta becomes optically thin is evaluated by equating the optical depth with unity.
Then, the condition that the ejecta is optically thick is written as
\begin{eqnarray}
t<t_{\rm{tr}}&\simeq&\sqrt{\frac{\xi\kappa{M}}{{\Omega}v^2}}=t_{\rm{diff}}\biggl(\frac{c}{v}\biggl)^{1/2},
   \label{t_tr 1}\\
&\simeq&12.4{\rm{\,day\,}}\kappa_{\rm{e},10}^{1/2}M_{\rm{e,0.01}}^{1/2}v_{{\rm{e,0.1}}}^{-1}\Omega_{0.5}^{-1/2}\biggl(\frac{\xi}{0.026}\biggl)^{1/2},
   \label{t_tr m}\\
&\simeq&2.0{\rm{\,day\,}}\kappa_{\rm{c},0.5}^{1/2}M_{\rm{c,0.005}}^{1/2}v_{{\rm{c,0.3}}}^{-1}\Omega_{0.5}^{-1/2}\biggl(\frac{\xi}{3/4\pi}\biggl)^{1/2}.
   \label{t_tr cocoon}
\end{eqnarray}
The second and third lines correspond to the timescales of the merger ejecta (with a density distribution) and the cocoon (in the one-zone approximation), respectively.
As long as $t_{\rm{diff}}<t<t_{\rm{tr}}$, the reprocessed luminosity is $L\sim{}L_{\rm{X}}$.
If the ejecta is otherwise diffusively thick $(t<t_{\rm{diff}})$, the reprocessed photons are trapped in the ejecta and suffer from adiabatic cooling.
If the ejecta is optically thin $(t>t_{\rm{tr}})$ instead, the reprocessed photons may not thermalize from X-ray to optical and NIR emissions.
We also remark that we use the grey opacity to estimate the optical depth for simplicity while the realistic opacity depends on wavelengths.

\begin{figure}
\begin{center}
\includegraphics[width=60mm, angle=270]{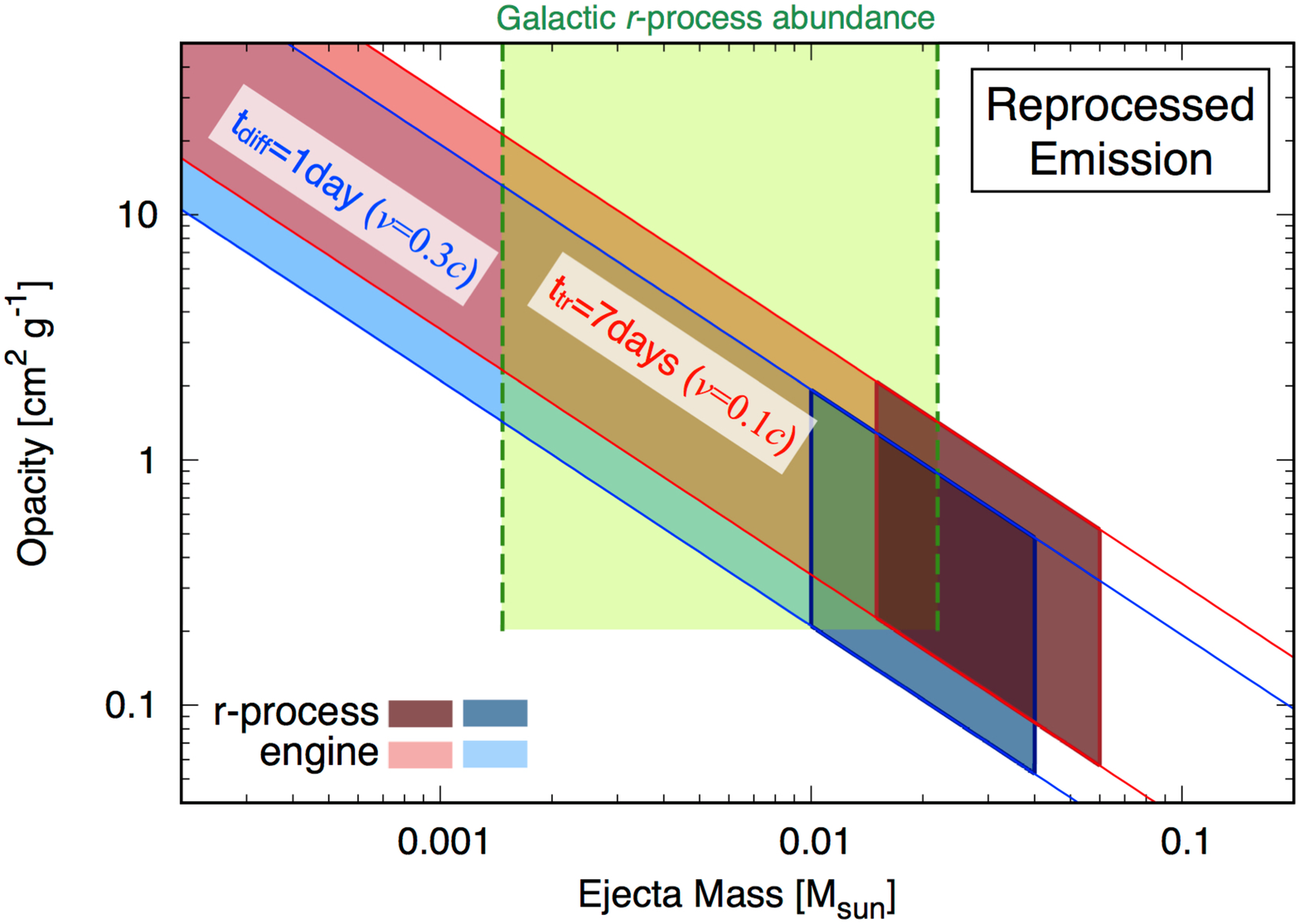}
\caption{The same as for Fig. \ref{fig km} but for the reprocessed emissions in the engine model.}
\label{fig km2}
\end{center}
\end{figure}

Since the reprocessed emission with a blackbody spectrum continues for $t_{\rm{diff}}<t<t_{\rm{tr}}$, we can constrain these timescales by observations.
When the blue macronova is produced by the reprocessed emission, its short timescale $\lesssim1\,\rm{day}$ gives an upper limit on the diffusion timescale, which results in a limit on $\kappa{M}$.
In Fig. \ref{fig km2}, we show this constraint with a blue shaded region.
This region is basically the same as the region in Fig. \ref{fig km} and Eq. \eqref{kappaM cool blue}.
On the other hand, since the red macronova showed spectra deviating from a blackbody at later time \citep[$\gtrsim7\,\rm{days}$,][]{Chornock+17,Kilpatrick+17,Nicholl+17,Pian+2017,Shappee+17}, these observations constraint the optically-thin timescale as $t_{\rm{tr}}\simeq7\,\rm{days}$, which results in a constraint on $\kappa{M}$ as
\begin{eqnarray}
\kappa_{\rm{e}}M_{\rm{e}}&\simeq&3.2\times10^{-2}\,{\rm{cm^2\,g^{-1}\,\Msun}}\nonumber\\
&&\biggl(\frac{v_{\rm{red}}}{0.1\,c}\biggl)^2\biggl(\frac{t_{\rm{tr}}}{7\,\rm{day}}\biggl)^2\biggl(\frac{\xi}{0.026}\biggl)^{-1}\Omega_{0.5}.
   \label{kappaM reprocess red}
\end{eqnarray}
We show this condition in Fig. \ref{fig km2} as a red shaded region.
When the ejecta satisfying this condition are irradiated by an X-ray excess, the reprocessed emission can explain the red macronova with $t_{\rm{tr}}\simeq7\,\rm{days}$.
Interestingly, both of the blue and red macronovae can be produced by the ejecta with a single combination of $\kappa{M}$ in the reprocessed emission model (but different velocities are required).

Note that the same condition (Eq. \eqref{kappaM reprocess red}) should be imposed on the ejecta in the \R model.
\cite{Waxman+2017} construct a light curve model taking the optically-thin timescale into account.
In particular, our results are consistent with their conclusions. 

\section{LIGHT CURVE MODELS}
\label{LIGHT CURVE MODEL}
Based on the order-of-magnitude estimation in the last section, we consider possible combinations of ejecta and energy injections, and construct simple light curve models in this section.
We show that the engine model can reproduce the observed macronova with various configurations.

\begin{table}[!t]
\begin{center}
\caption{Possible combinations of ejecta and energy injections}
\label{table combination}
\begin{tabular}{rc|c||c|c}
\tableline\tableline
&\multicolumn{2}{c||}{Blue Macronova} & \multicolumn{2}{c}{Red Macronova}\\
\tableline
&ejecta&energy&ejecta&energy\\
&($\kappa{M}\sim10^{-3}$)&source&($\kappa{M}\sim10^{-2}$)&source\\
\tableline\tableline
(A)&cocoon&jet&merger ejecta(eq)&X-ray\\
(B)&cocoon&X-ray&merger ejecta(eq)&X-ray\\
(C)&merger ejecta(po)&X-ray&merger ejecta(eq)&X-ray\\
(D)&merger ejecta(eq)&X-ray&merger ejecta(po)&X-ray\\
\end{tabular}
\end{center}
{{\bf{Notes.}} The values of $\kappa{M}$ are written in a unit of $\rm{cm^2\,g^{-1}\,\Msun}$. The location (polar and equatorial directions) of the merger ejecta is represented by ``po" and ``eq", respectively.}
\end{table}

\begin{figure}
\begin{center}
\includegraphics[width=60mm, angle=270]{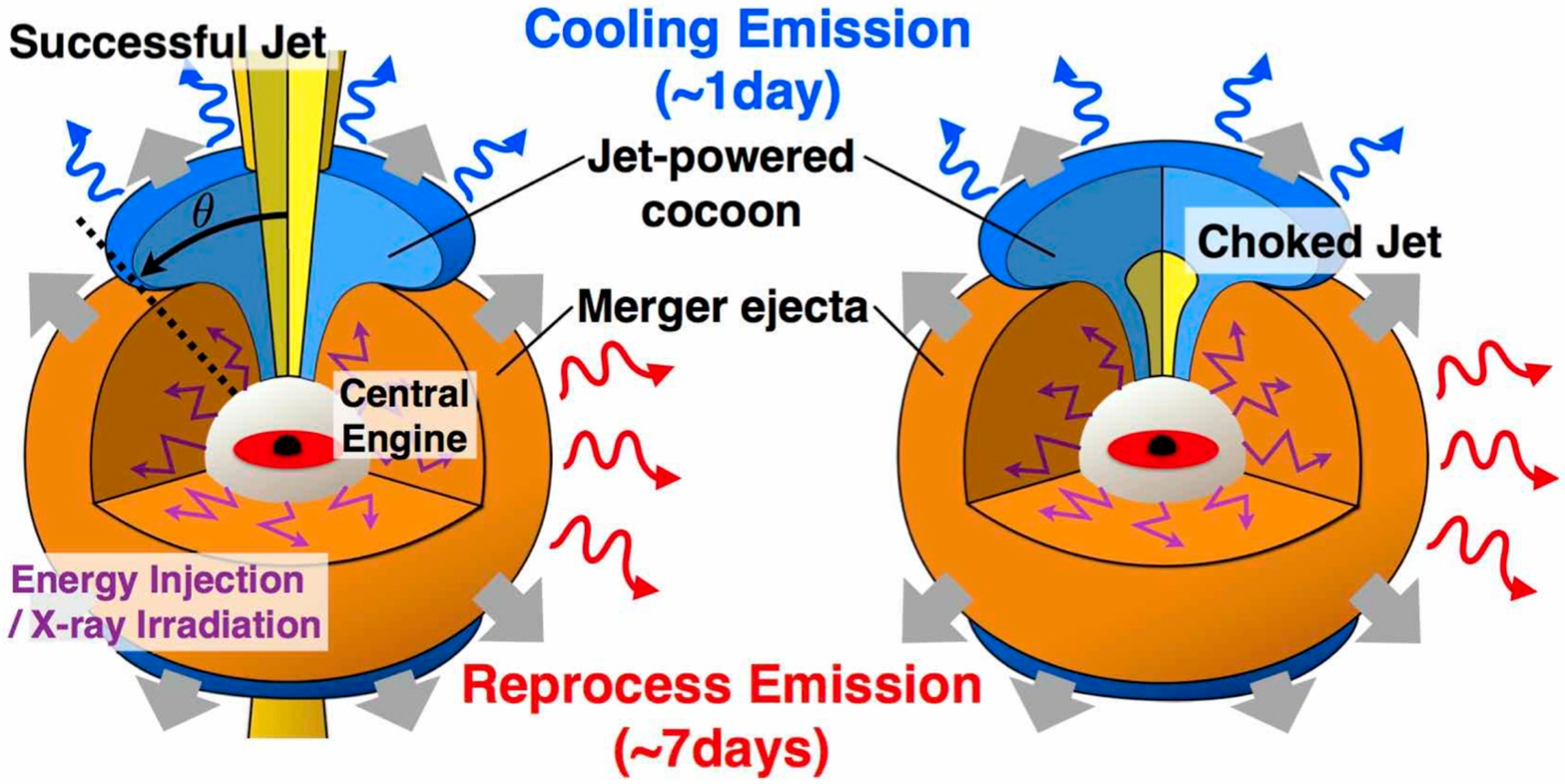}
\caption{The schematic picture of the jet-powered cocoon and the merger ejecta, which are powered by the jet activities and the quasi-isotropic X-ray excess, respectively.
\textit{Left} and \textit{right} figures show the cases of the successful and the choked jets. These figures correspond to model (A) in Table \ref{table combination}.}
\label{fig picture}
\end{center}
\end{figure}

In Table \ref{table combination}, we list the possible combinations of ejecta and energy sources to reproduce the macronova.
Since the engine model allows the large parameter space (see Figs. \ref{fig km} and \ref{fig km2}), various combinations are possible.
When we consider two component ejecta such as the cocoon and the merger ejecta, we have two situations depending on their energy sources.
In these models, we consider the merger ejecta in the equatorial direction, which have the large opacity and produce a long emission timescale.
In model (A), the cocoon and the merger ejecta are separately powered by the jet activity and X-ray excess emission.
Then, the diffusion emission from the cocoon and the reprocessed emission from the merger ejecta produce the blue and red macronovae, respectively.
In Fig. \ref{fig picture}, we show schematic pictures of this model.
As we discuss in section \ref{Cocoons}, regardless of the successful and choked jets (left and right panels of Fig. \ref{fig picture}), the cocoon may break out of the merger ejecta.
In model (B), both cocoon and merger ejecta are powered by the X-ray excess and produce the blue and red macronovae, respectively.
In this model, we assume that the jet activity is not so powerful to power the cocoon.

We can also consider the cases where both blue and red macronovae are produced by the merger ejecta.
Since the merger ejecta (dynamical ejecta) have the directional opacity distribution, the polar and equatorial directed ejecta may produce the blue and red macronovae, respectively, if they are irradiated by the X-ray excess.
Even with a constant opacity, if the ejecta mass has a polar distribution, the polar ejecta with a small mass and the equatorial ejecta with a large mass can produce the blue and red macronovae, respectively.
We name these situations as model (C).
As another possibility, when the post-merger ejecta (polar directed) have larger mass and hence larger $\kappa{M}$ than the dynamical ejecta (equatorial directed), the merger ejecta in the polar and equatorial directions may produce the red and blue macronovae, respectively, irradiated by the X-ray luminosity (model (D)). 

In the reprocessed emission model, we do not consider the configuration where a single component ejecta has a radial $\kappa{M}$ distribution, because such profile can not reproduce the emission timescales.
Since X-rays are absorbed at the inner part of the ejecta due to a large $\kappa{M}$, the reprocessed photons can not leak out from this inner part in the observed short timescale. 

In the following subsections, we construct simple light curve models focusing on models (A) and (B).
The same procedure is applied to the other models and gives similar light curves.

\subsection{Cooling Emission from Jet-Powered Cocoons}\label{Diffusion Emission from Jet-Powered Cocoons}
First, we consider the light curve of the cooling emission from the jet-powered cocoon, which produces the blue macronova in model (A).
We construct a simple emission model by a one-zone approach.
The jet-powered cocoon with mass $M_{\rm{c}}$ and velocity $v_{\rm{c}}$ expands homologously into a fraction $\Omega$ of the solid angle.
The outermost radius of the ejecta is given by $R=v_{\rm{c}}t$, which coincides with the photosphere for the early phase of $t\lesssim{t_{\rm{diff}}}(<t_{\rm{tr}})$. 
The time evolution of the diffusion luminosity is simply given by considering the thermodynamics of the ejecta \citep{Arnett1980,2016ApJ...823...83M}.
The first law of thermodynamics is given by
\begin{eqnarray}
\frac{dE}{dt}=-P\frac{dV}{dt}-L+H,
\label{ODE}
\end{eqnarray}
where $E$, $P$, $V=4\pi\Omega{R^3}/3$, $L$, and $H$ is the total internal energy, pressure, volume, diffusion luminosity, and heating from the radioactive decay.
The pressure is dominated by the radiation $P=E/3V$.
The diffusion luminosity is evaluated by the diffusion approximation as $L\simeq{tE}/t_{\rm{diff}}^2$.
We neglect the radioactive decay heating, which is justified as long as the cocoon mass is smaller than that required in the \R model ($\simeq0.02\,\Msun$, see Fig. \ref{fig km}).
With the initial condition of $E=E_{\rm{in}}$ at $t=t_{\rm{in}}$, we can easily integrate Eq. \eqref{ODE} and obtain the time evolution of the internal energy and the luminosity as
\begin{eqnarray}
L(t)=\frac{E_{\rm{in}}t_{\rm{in}}}{t_{\rm{diff}}^2}\exp\biggl(-\frac{t^2-t_{\rm{in}}^2}{2t_{\rm{diff}}^2}\biggl).
\end{eqnarray}
The observed photospheric temperature is given by using the photospheric radius $R_{\rm{ph}}\sim{v_{\rm{c}}t}$ and the bolometric luminosity as
\begin{eqnarray}
T_{\rm{ph}}=\biggl(\frac{L}{4\pi\Omega{R_{\rm{ph}}^2}\sigma}\biggl)^{1/4}.
\label{T_obs}
\end{eqnarray}
We also assume the blackbody spectrum to depict light curves \citep{Cowperthwaite+17,Kilpatrick+17,McCully+17,Nicholl+17,Shappee+17,Valenti+17}.

\begin{figure}
\begin{center}
\includegraphics[width=60mm, angle=270]{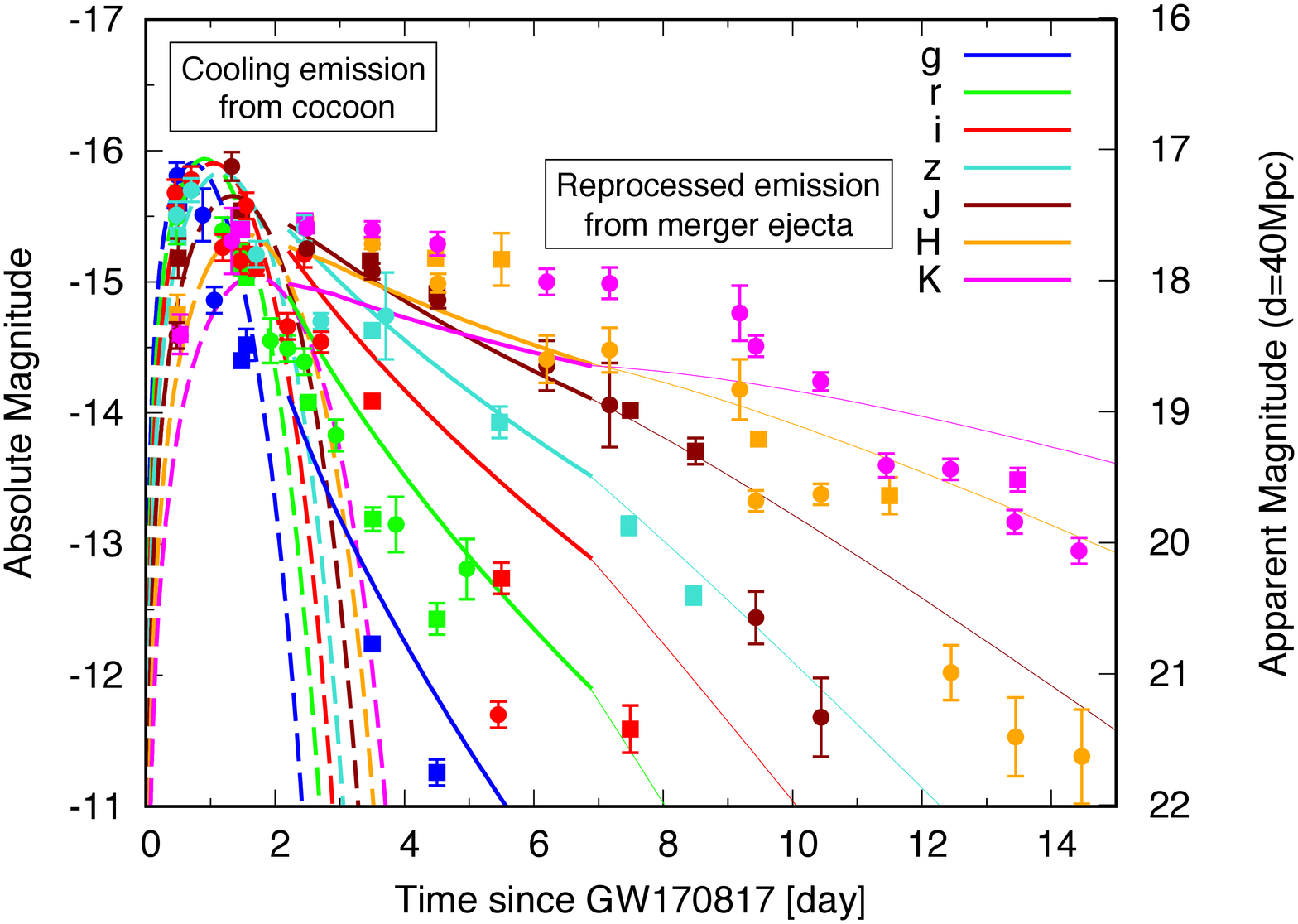}
\caption{The light curves of the diffusion emission from jet-powered cocoon (dashed) and the reprocessed emission from the merger ejecta (solid), respectively (model (A)).
For the merger ejecta, we show the light curves in the diffusively thin and optically thick phase $t_{\rm{diff}}<t<t_{\rm{tr}}$ and optically thin phase $t_{\rm{tr}}<t$ with thick and thin solid curves, respectively.
Both light curves show good agreements with the observed data points taken from \cite{Kasliwal+2017} (circle) and \cite{Drout+2017} (square).
By the jet activities, the jet-powered cocoon receives the energy which powers the blue macronova.
The red macronova is emitted from the merger ejecta as the NIR emission reprocessed from the quasi-isotropic X-ray excess produced by the central engine.}
\label{fig lc}
\end{center}
\end{figure}

In Fig. \ref{fig lc}, the light curves of the jet-powered cocoon show a good agreement with the observed blue macronova.
We show the light curves with dashed curves, and the observed data points taken from \cite{Kasliwal+2017} and \cite{Drout+2017}.
The parameters of the ejecta and energy injections are listed in Tables \ref{table ejecta}.
The required injection energy (or $E_{\rm{in}}t_{\rm{in}}$) are supplied by jet activities, such as the prompt, extended, and plateau emissions.
The light curves show a rapid decay after $t\gtrsim{}t_{\rm{diff}}$ because the photons diffuse out after the diffusion time.

\begin{figure}
\begin{center}
\includegraphics[width=60mm, angle=270]{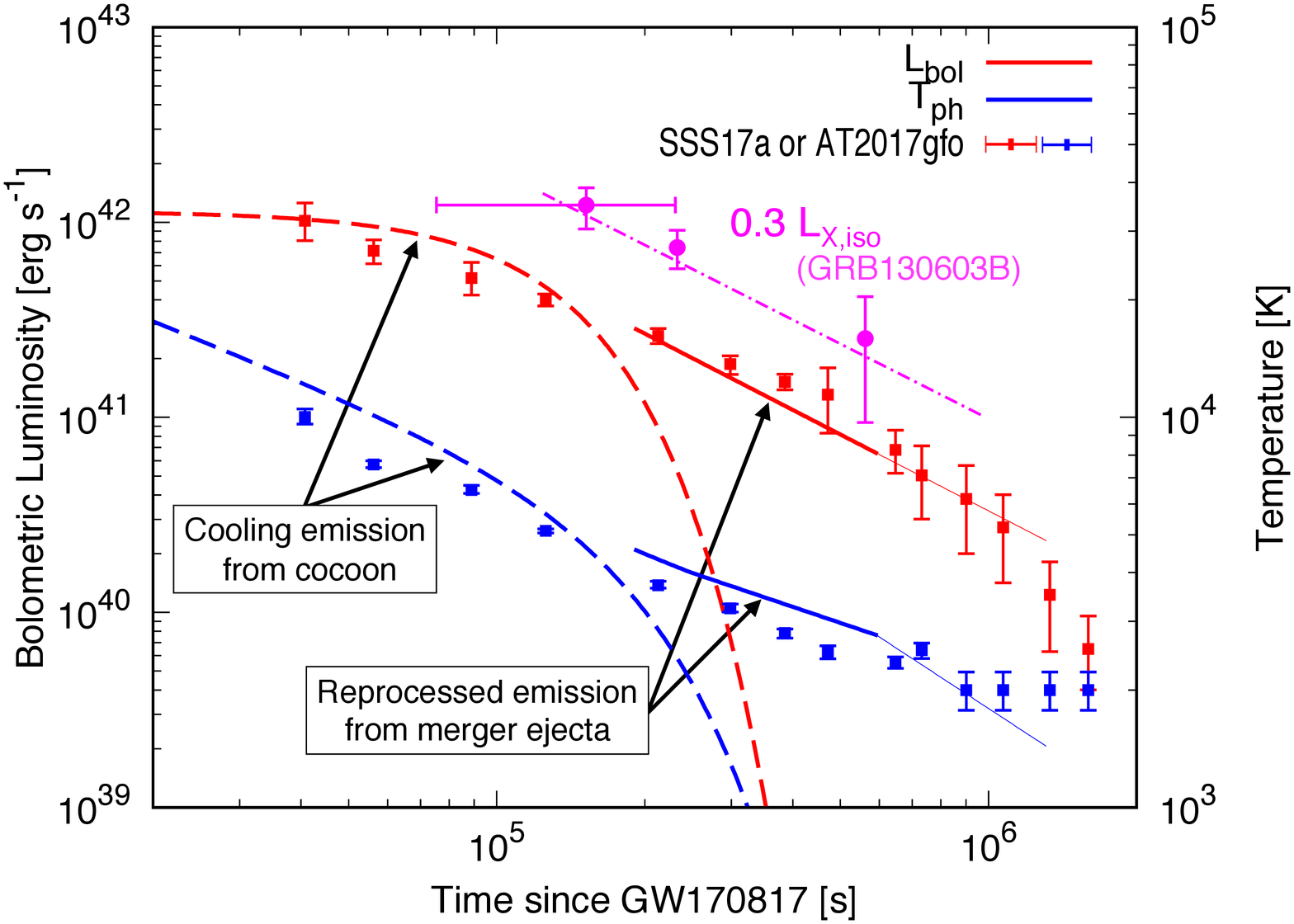}
\caption{The time evolution of bolometric luminosity (red) and photospheric temperature (blue) of the jet-powered cocoon (dashed; blue macronova) and the merger ejecta (solid; red macronova), respectively, in model (A).
The red and blue data points denote the observed luminosity and temperature of the red and blue macronovae (SSS17a or AT2017gfo), which are taken from \cite{Kilpatrick+17}.
The engine model successfully reproduces the observation data.
We also show the observed X-ray excess luminosity of GRB 130603B taken from \cite{Fong+2014}, which has the same slope with the red macronova, $\alpha_{\rm{X}}\simeq1.3$.}
\label{fig lt}
\end{center}
\end{figure}

In Fig. \ref{fig lt}, the red and blue dashed curves denote the bolometric luminosity and the photospheric temperature, respectively.
We also show the observed bolometric luminosity, which is derived by summing up the flux in each band, and the photospheric temperature, which fits the observed spectra with a blackbody \citep{Kilpatrick+17}.
The calculated luminosity and temperature roughly agree with the observed ones.

\cite{Piro&Kollmeier2017} also consider the early-optical emission as the cocoon emission and construct a light curve model based on \cite{2017ApJ...834...28N}.
In contrast to our one-zone approach, they assume an internal structure of the ejecta and calculate the light curve.
They mainly focus on the prompt jet as the energy injection channel, while we consider more injection processes such as the extended and plateau emissions.

\begin{table*}[!t]
\begin{center}
\caption{The fiducial parameters of the cocoon and the merger ejecta for model (A) in Table \ref{table combination}.}
\label{table ejecta}
\begin{tabular}{lccc}
\tableline\tableline
&Symbol & Cocoon & Merger Ejecta \\
\tableline
Opacity times Mass &$\kappa{M}$ & $0.0025\rm{\,cm^2\,g^{-1}}\,\Msun$ & $0.03\rm{\,cm^2\,g^{-1}}\,\Msun$ \\
Velocity &$v_{\rm{c}}$, $v_{\rm{max/min}}$ & $0.3\,c$ & $0.1-0.4\,c$ \\
Subtended Solid Angle & $\Omega$ & 0.5 & 0.5 \\
Power Law Index of Density Profile & $\beta$ & - (one-zone) & 3.5 \\
\tableline
Injection Energy times Time & $E_{\rm{in}}t_{\rm{in}}$ &$ 10^{52}\,\rm{erg\,s}$ & - \\
X-ray Excess Luminosity & $L_{\rm{X}}(t)$ & - & $8\times10^{41}\,(t/\rm{day})^{-1.3}\rm{\,erg\,s^{-1}}$\\
\end{tabular}
\end{center}
{{\bf{Notes.}} The ejecta mass should be smaller than that required in the \R model ($M_{\rm{c}}\lesssim0.02\,\Msun$ and $M_{\rm{e}}\lesssim0.03\,\Msun$) in order to neglect the \R heating.}
\end{table*}

\subsection{Reprocessed Emission}\label{reprocessed emission from merger ejecta}
Next, we consider the reprocessed emissions from the merger ejecta (for models (A)-(D)) and the cocoon (for model (B)).

\subsubsection{Reprocessed Emission from Merger Ejecta}
We first discuss the emission from the merger ejecta using the formulation in \cite{Kisaka+2015}, and later apply the same formalism to the cocoon.
Since the energy injection from the central engine is not significant, the merger ejecta travel homologously keeping the original density profile.
The maximum and minimum velocities (corresponding to the velocities at the outermost and innermost radii of the ejecta) are denoted as $v_{\rm{max}}$ and $v_{\rm{min}}$, respectively.
Here, we assume that the ejecta is mainly composed of the dynamical ejecta with a power-law density profile described by
\begin{eqnarray}
\rho(r,t)=\frac{f(\beta,v_{\rm{max}}/v_{\rm{min}})}{4\pi\Omega}\frac{M_{\rm{e}}}{R_{\rm{in}}^3}\biggl(\frac{r}{R_{\rm{in}}}\biggl)^{-\beta},
\end{eqnarray}
where $f(\beta,v_{\rm{max}}/v_{\rm{min}})=(\beta-3)/\bigl[1-(\frac{v_{\rm{max}}}{v_{\rm{min}}})^{3-\beta}\bigl]$ is the normalization factor of the density, and $R_{\rm{in}}(=v_{\rm{min}}t)$ is the inner radius of the ejecta.
Numerical simulations suggest that the index $\beta$ has a range of $3\lesssim\beta\lesssim4$ \citep{Hotokezaka+2013,2014ApJ...784L..28N}. 

We define two characteristic radii.
One is the diffusion radius $R_{\rm{diff}}$, where the dynamical time is equal to the diffusion time, and the photospheric radius $R_{\rm{ph}}$, where the optical depth $\tau$ is unity.
The optical depth at a radius $r$ is evaluated by\footnote{Since we mainly focus on the diffusively thin phase, we do not consider the early time evolution of the radii, where the radii $2R_{\rm{diff}}$ and $2R_{\rm{ph}}$ are larger than the outer radius $R_{\rm{out}}(=v_{\rm{max}}t)$, so-called the thin diffusion phase \citep{Kisaka+2015}.}
\begin{eqnarray}
\tau(r)=\int_{r}^{2r}\kappa_{\rm{e}}\rho{dr^\prime}=\frac{f\kappa_{\rm{e}}M_{\rm{e}}}{4\pi\Omega{R_{\rm{in}}}^2}\frac{1-2^{1-\beta}}{\beta-1}\biggl(\frac{r}{R_{\rm{in}}}\biggl)^{1-\beta}.
\label{tau thick}
\end{eqnarray}
The diffusion radius at time $t$ is given by
\begin{eqnarray}
R_{\rm{diff}}=\frac{c}{\tau(R_{\rm{diff}})}{t}.\label{t_diff}
\label{diffusion time}
\end{eqnarray}
Then, we obtain the diffusion radius by substituting Eq. \eqref{tau thick} into Eq. \eqref{diffusion time},
\begin{eqnarray}
R_{\rm{diff}}=R_{\rm{in}}\biggl[\frac{f(1-2^{1-\beta})}{\beta-1}\frac{\kappa_{\rm{e}}{M_{\rm{e}}}}{{4\pi}v_{\rm{min}}ct^2}\biggl]^{\frac{1}{\beta-2}}\label{r_diff}\propto{t^{\frac{\beta-4}{\beta-2}}}.
\end{eqnarray}
The photosphere is given by equating Eq. \eqref{tau thick} with unity, 
\begin{eqnarray}
R_{\rm{ph}}=R_{\rm{in}}\biggl[\frac{f(1-2^{1-\beta})}{\beta-1}\frac{\kappa_{\rm{e}}{M_{\rm{e}}}}{{4\pi\Omega}v_{\rm{min}}^2t^2}\biggl]^{\frac{1}{\beta-1}}\label{r_diff}\propto{t^{\frac{\beta-3}{\beta-1}}}.
\label{r_ph}
\end{eqnarray}

We estimate the times $t_{\rm{diff}}$ and $t_{\rm{tr}}$ when the diffusion and photospheric radii reach the innermost ejecta radius, respectively.
Each timescale corresponds to the diffusion time and the transparent timescale discussed in section \ref{engine model}.
By equating the diffusion radius and the photosphere with the innermost radius of ejecta, we get
\begin{eqnarray}
t_{\rm{diff}}&=&\sqrt{\frac{f(1-2^{1-\beta})}{\beta-1}\frac{\kappa_{\rm{e}}{M_{\rm{e}}}}{{4\pi\Omega}v_{\rm{min}}c}},\label{t_diff}\\
t_{\rm{tr}}&=&\sqrt{\frac{f(1-2^{1-\beta})}{\beta-1}\frac{\kappa_{\rm{e}}{M_{\rm{e}}}}{{4\pi\Omega}v_{\rm{min}}^2}},\label{t_tr}
\end{eqnarray}
respectively.
As we discussed in section \ref{engine model}, the density structure gives the coefficient $\xi=f(1-2^{1-\beta})/(\beta-1)$, which reduce the timescales about $\sim0.3$ from the values given by the one-zone estimation.

When the merger ejecta is diffusively thin and optically thick, $t_{\rm{diff}}<t<t_{\rm{tr}}$, X-rays radiated from the central engine are absorbed in the ejecta, and reprocessed to lower energy photons with a blackbody spectrum.
We should remind that when the ejecta is otherwise diffusively thick, the reprocessed photons can not leak out of the ejecta.
Furthermore, it should be noted that the reprocessed photons do not show the thermal spectrum if the ejecta is optically thin.
As we discussed in section \ref{Long-Lasting X-ray Luminosity}, from the observation that the red macronova shows spectra deviating from a blackbody at $\sim7\,\rm{days}$, the ejecta parameter $\kappa_{\rm{e}}{M_{\rm{e}}}$ is constrained as given Eq. \eqref{kappaM reprocess red}.
This condition is satisfied with reasonable merger ejecta parameters such as $\kappa_{\rm{e}}\simeq10\,\rm{cm^2\,g^{-1}}$ and $M_{\rm{e}}\simeq3\times10^{-3}\,\Msun$ (see also section \ref{Merger Ejecta}). 
The bolometric light curve of the reprocessed emission follows the irradiation luminosity $L_{\rm{bol}}(t)=L_{\rm{X}}(t)$.
Motivated by the X-ray excess detected in GRB 130603B with power-law temporal decay \citep{Fong+2014}, we also assume a power-law-decaying irradiation luminosity of
\begin{eqnarray}
L_{\rm{X}}(t)=L_{\rm{X}}\biggl(\frac{t}{\rm{day}}\biggl)^{-\alpha_{\rm{X}}},
   \label{irradiation luminosity}
\end{eqnarray}
where $L_{\rm{X}}$ is the normalization.
The power-law decay might be related with the fallback accretion, whose mass accretion rate would also show the power-law temporal decay with index $\sim5/3$.
We determine the photospheric temperature with Eqs. \eqref{T_obs} and \eqref{r_ph}. 

In Fig. \ref{fig lc}, we show the light curves of the reprocessed emission from the merger ejecta with solid curves.
In Table \ref{table ejecta} , we show the adopted parameters.
The adopted X-ray irradiation luminosity is consistent with the X-ray observations such as \textit{Chandra} and \textit{NuSTAR} (see, section \ref{DISCUSSION}).
The reprocessed emission also shows good agreements with the observed red macronova.
We extend the light curves after the merger ejecta becomes optically thin ($t>t_{\rm{tr}}$) with thin solid curves, by assuming $R_{\rm{ph}}=R_{\rm{in}}$ and a blackbody spectrum.

In Fig. \ref{fig lt}, we show the bolometric luminosity and temperature with solid red and blue lines, respectively.
The magenta data points and dash-dotted line are the isotropic X-ray excess luminosity observed in GRB 130603B taken from \cite{Fong+2014}, which we convert from the observed flux.
We also reduce the X-ray luminosity by multiplying 0.3 to compare with the macronova's bolometric luminosity.
The luminosity of the X-ray excess in GRB 130603B has a similar slope to that of the macronova.
Note that the temporal index $\alpha_{\rm{X}}$ which we adopt (see Table \ref{table ejecta}) is roughly similar to the index of the fallback accretion, $5/3\simeq1.67$, and the index of the X-ray excess in GRB 130603B, $\simeq1.3$ \citep{Fong+2014}.
\cite{Smartt+2017} show that the light curve is fitted by choosing the temporal index $\sim1.2\pm0.3$, which is favored by the \R heating.
While they conclude that SSS17a may be powered by the radioactive decay heating by \R elements, our result suggests another possibility, i.e., the engine-powered macronova.

As shown in Fig. \ref{fig lc}, the reprocessed emission lasts for $t_{\rm{diff}}<t<t_{\rm{tr}}$ and the light curves do not connect with the light curves of the jet-powered cocoon.
This is natural because we consider only two component (two combinations of $\kappa{M}$) discrete ejecta.
In the realistic situation, there should be a gradient of the opacity ($0.1 \lesssim\kappa/\rm{cm^{2}\,g^{-1}}\lesssim10$) or the ejecta mass, and fill the gap in the both light curves smoothly.
Furthermore, the reprocessed emission rises at $t\lesssim{t_{\rm{diff}}}$, and this also fills up the gap.

\subsubsection{Reprocessed Emission from Cocoons}
Finally, we discuss the reprocessed emission from the cocoon when the cocoon does not receive a significant energy injection from jet activities (model (B)).
Here, we simply calculate the emission by using a one-zone model as in section \ref{Diffusion Emission from Jet-Powered Cocoons}.
We derive the temporal evolution of the photospheric radius by solving,
\begin{eqnarray}
1=\tau=\kappa\rho\Delta{R},
\end{eqnarray}
where $\Delta{R}=R-R_{\rm{ph}}$.
The photospheric radius is obtained as 
\begin{eqnarray}
R_{\rm{ph}}=R\biggl(1-\frac{t^2}{t_{\rm{tr}}^2}\biggl),
\end{eqnarray}
which shows that the photosphere shrinks at later time $t\sim{t_{\rm{tr}}}$.
However, the cocoon may have a complex density structure in the innermost region due to the mixing or shock interactions with the merger ejecta.
In particular, before the photosphere reaches the innermost radius of the cocoon ($R_{\rm{ph}}\to0$), the one-zone approximation breaks down.
Therefore, we assume that the innermost radius of the cocoon is roughly equal with that of the merger ejecta $R_{\rm{in}}$.
We also depict the light curve for $t<t_{\rm{diff}}$ and $t\gtrsim{t_{\rm{tr}}}$ by suppressing the irradiation luminosity with exponential cut-offs.

\begin{figure}
\begin{center}
\includegraphics[width=60mm, angle=270]{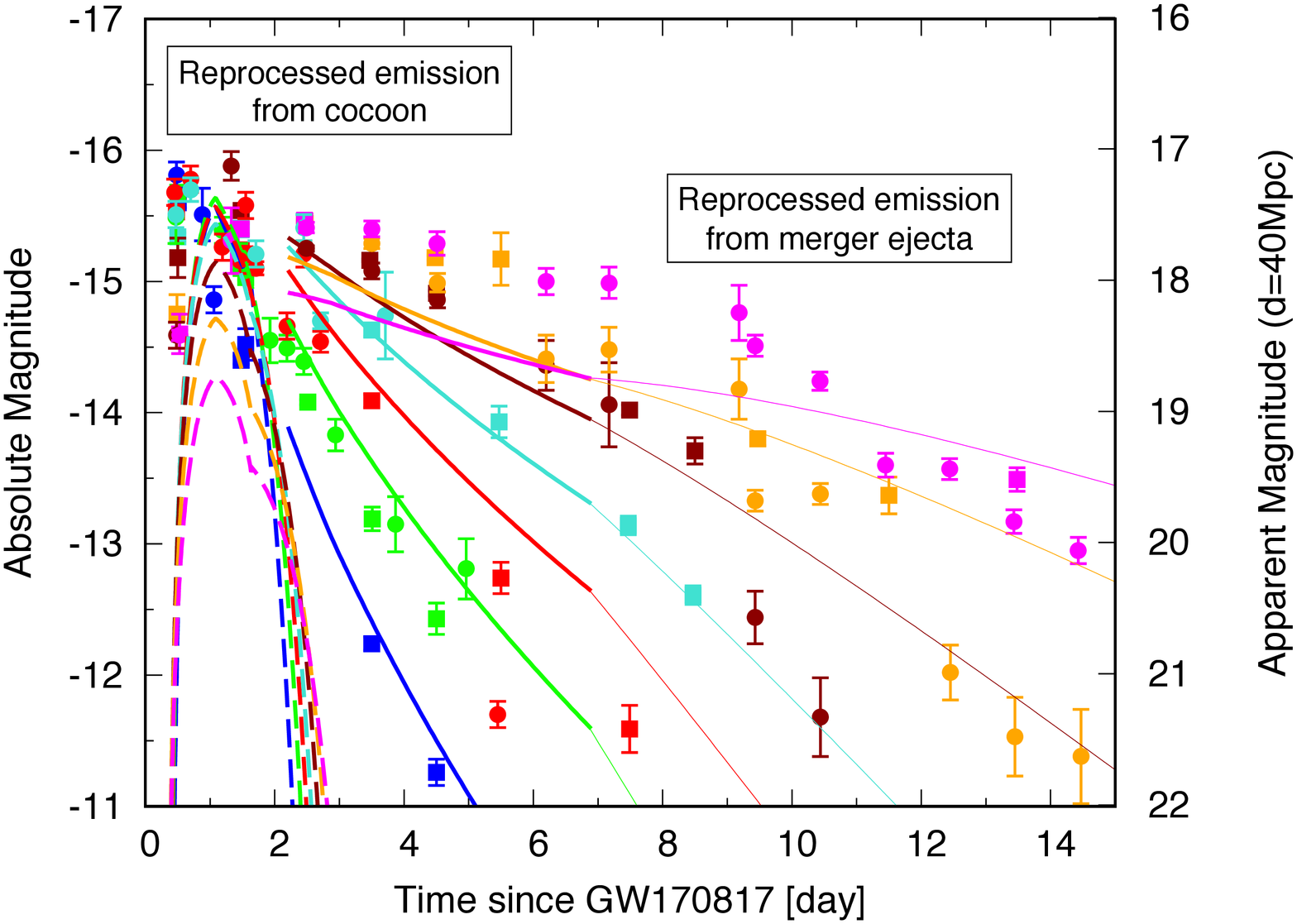}
\caption{The same as Fig. \ref{fig lc} but for model (B).
Both blue and red macronovae are powered by the reprocessed emissions from the cocoon (solid) and the merger ejecta (dashed), respectively.}
\label{fig lc2}
\end{center}
\end{figure}

In Fig. \ref{fig lc2}, we show the light curves of the reprocessed emission from the cocoon in addition to those of the merger ejecta (model (B)).
Both light curves are drawn by using a single X-ray excess luminosity with the same function form of Eq. \eqref{irradiation luminosity}.
We adopt the same ejecta parameters as in Fig. \ref{fig lc}, but the X-ray luminosity of $L_{\rm{X}}(t)=7\times10^{41}\,(t/\rm{day})^{-1.3}\rm{\,erg\,s^{-1}}$.
The cocoon parameters are the same as the values in Table \ref{table ejecta}, which satisfy the observational constraint of $t_{\rm{diff}}\lesssim1\,\rm{day}$ (see also Eq. \eqref{kappaM cool blue}).
The reprocessed emission from the cocoon has a rather higher temperature than the observed one because in the one-zone model, the photosphere just recedes, which raises the photospheric temperature.
In order to obtain more detailed temporal behavior of the photosphere, we should consider the density structure of the cocoon.
It should be noted again that the gap between the light curves of the cocoon and the merger ejecta is filled up by the emission from the ejecta with $\kappa{M}$ connecting the value of the cocoon and the merger ejecta (section \ref{reprocessed emission from merger ejecta}).

\section{DISCUSSION}\label{DISCUSSION}
In this work, we study whether the energy injection from the central engine rather than the \textit{r}-process radioactive decay can produce the observed blue and red macronova emissions or not.
Since in the engine model, only the product of the ejecta mass and opacity $\kappa{M}$ is constrained by the observed emission timescale, large parameter spaces are allowed (see Figs. \ref{fig km} and \ref{fig km2}).
The observations suggest that the ejecta has a polar or radial distribution of $\kappa{M}$ from $\kappa{M}\simeq2\times10^{-3}{\,\rm{cm^2\,g^{-1}}\,\Msun}$ (with $v\simeq0.3\,c$ for the blue macronova at 1 day, see Eq. \eqref{kappaM cool blue}) to $\kappa{M}\simeq3\times10^{-2}{\,\rm{cm^2\,g^{-1}}\,\Msun}$ (with $v\simeq0.1\,c$ for the red macronova at 1 week, see Eq, \eqref{kappaM reprocess red}).
We can consider various possible configurations of ejecta and energy sources (see Table \ref{table combination}).
In particular, we depict light curves based on two models in section \ref{LIGHT CURVE MODEL}.
In model (A), we consider jet activities and X-rays from the matter fallback as energy sources.
While the jet activities inject energy into a cocoon, the X-rays are absorbed by merger ejecta and reprocessed into NIR photons.
In model (B), both the cocoon and merger ejecta are irradiated by X-rays to power reprocessed emissions.
With reasonable amounts of the injection energy, the diffusion or reprocessed emission from the cocoon and the reprocessed emission from the merger ejecta reproduce the observed blue and red macronovae, respectively.
The reprocessed emission of the X-rays is motivated by the observed X-ray excess in GRB 130603B, which can explain the luminosity and duration of the NIR macronova associated with GRB 130603B with a single energy source, i.e., the central engine.
With the polar distributions of the ejecta mass or opacity, even a single-component ejecta can reproduce the observed blue and red macronovae in the engine model (models (C) and (D)).

The necessary mass in the engine model can be smaller than that required by the \textit{r}-process model (see Fig. \ref{fig km} and \ref{fig km2}).
As long as the products $\kappa{M}$ are fixed, we can adopt various values of opacity and ejecta mass.
In particular, as shown in Figs. \ref{fig lc} and \ref{fig lt}, the engine model can reproduce the observed blue and red macronovae with the ejecta mass of $M_{\rm{ej}}^{\rm{blue}}=M_{\rm{c}}\simeq0.005\,\Msun\,(\kappa_{\rm{c}}/0.5\,\rm{cm^2\,g^{-1}})^{-1}$ and $M_{\rm{ej}}^{\rm{red}}=M_{\rm{e}}\simeq0.003\,\Msun\,(\kappa_{\rm{e}}/10\,\rm{cm^2\,g^{-1}})^{-1}$, respectively.
On the other hand, the \textit{r}-process model requires more mass of $M_{\rm{ej}}^{\rm{blue}}\,,M_{\rm{ej}}^{\rm{red}}\gtrsim0.02-0.03\,\Msun$.
The smaller ejecta mass than that in the \R model can resolve the concerns which we raise in section \ref{intro}.

Although we illustrate the above ejecta mass and opacity values motivated by numerical simulations, the different values can be also allowed.
Then, we can discuss the minimum required mass or opacity by fixing the other parameter values.
For the jet-powered cocoon, if the ejecta have significant \R elements $\kappa_{\rm{c}}\simeq10\,\rm{cm^2\,g^{-1}}$, only a small amount of ejecta mass $M_{\rm{c}}\simeq3\times10^{-4}\,\Msun$ is sufficient to power the blue macronova.
In this case, the injection energy should be smaller than the ejecta' kinetic energy $\sim10^{49}\,\rm{erg}$.
Then, since the required injection energy is $E_{\rm{in}}t_{\rm{in}}=10^{52}\,\rm{erg\,s}$ (see Table \ref{table ejecta}), only plateau emissions ($t_{\rm{in}}>10^3\,\rm{s}$) could reproduce the observed blue macronova.
If future radio observations will give a constraint on the ejecta mass ($\lesssim10^{-3}\,\Msun$), the blue macronova strongly supports the existence of the long-timescale engine activity in sGRB 170817A.

As we discuss in the previous sections, while the \R elements may not be essential as an energy source, they are likely an opacity source.
By considering a reasonable range of the merger ejecta mass, we can show that the \R elements may be necessary to explain the observed emission timescale.
Regardless of the energy source, e.g., the \R heating or the central engine, photons should be thermalized to the NIR energy in the ejecta.
Therefore, at least a part of the ejecta should be optically thick to NIR photons, and the emission time has to satisfy the condition, $t<t_{\rm{tr}}\simeq12.4\,{\rm{day}}\,\kappa_{\rm{e,10}}^{1/2}M_{\rm{e,0.01}}^{1/2}v_{\rm{e,0.1}}^{-1}\Omega_{0.5}^{-1/2}(\xi/0.026)^{1/2}$.
Even for a large mass ejecta of $M_{\rm{e}}=0.1\,\Msun$, this condition requires the large opacity as 
\begin{eqnarray}
\kappa_{\rm{e}}&>&0.31\,{\rm{cm^{2}\,g^{-1}}}\,\biggl(\frac{t}{7\,{\rm{day}}}\biggl)^{2}\nonumber\\
&&\biggl(\frac{v_{\rm{min}}}{0.1\,c}\biggl)^{2}\biggl(\frac{M_{\rm{e}}}{0.1\,\Msun}\biggl)^{-1}\Omega_{0.5}\biggl(\frac{\xi}{0.026}\biggl)^{-1},
\end{eqnarray}
where the velocity $v_{\rm{min}}$ is not able to be changed a lot.
While this constraint is not so strong as to require \R elements definitely, it is easily satisfied by the small amount of \R elements \citep[Lanthanoids,][]{Kasen+2013,Tanaka&Hotokezaka2013,Tanaka+2017}.
Thus, the strongest evidence of the \R elements so far is the long duration of the red macronova emission, neither the temporal index of the bolometric light curve, which may be reproduced by the energy injection from the X-ray excess, nor the spectral lines, for which there remain theoretical uncertainties.
We remark that other than the \R elements, dust grains are also proposed as an opacity source \citep{Takami+2014}.
However, the observed line feature in the spectrum \citep{Chornock+17,Kilpatrick+17,Nicholl+17,Pian+2017,Shappee+17} may not prefer this possibility \citep[see also,][]{Gall+2017}.

The merger ejecta satisfying Eq. \eqref{kappaM reprocess red} becomes optically thin $\sim7\,\rm{days}$ after the merger $(t>t_{\rm{tr}})$.
In this case, the reprocessed photons are not thermalize completely, and may show deviations from the thermal spectrum in bluer wavelength.
In the observations of SSS17a, the spectra actually deviate from a blackbody at late time.
While these spectra are roughly fitted in the \R model \citep{Chornock+17,Kilpatrick+17,Nicholl+17,Pian+2017,Shappee+17}, we could also fit them in the engine model.
In order to predict the non-thermalized spectrum, we need a detailed radiative transfer calculation, which is beyond the scope of this paper.
This is an interesting future problem to compare the spectra given by both models and clarify whether macronovae are produced by the energy injection from the central engine or not.

In addition to late-time spectra, light curves at the early phase may be useful to study what powers macronovae.
While the \R heating rate is well understood, various energy injection processes are possible in the engine model.
Therefore, the difference of the heating mechanisms may be reflected in the early-time light curves \citep[$t\lesssim1\,\rm{day}$, see e.g., Fig. 6 in][]{Kisaka+2015}.

The observed macronova shows a smooth bolometric light curve for $\sim1-7\,\rm{days}$ \citep{Smartt+2017,Waxman+2017}.
Thus, the single-energy-source scenario (models (B)-(D)) can more naturally reproduce the observed light curve than the double-energy-source scenario (model (A)).
Although reasonable jet parameters gives the diffusion luminosity comparable to the reprocessed emission in model (A), the possible injected energy by jets has a relatively broad range.
Then, the detection of very bright blue macronovae which requires too much mass to explain the event in the \R model, can be a smoking gun of the engine model (for the case of model (A)).

We also discuss the X-ray detection by \textit{Chandra} and \textit{NuSTAR} \citep{Troja+2017,Ruan+2017,Evans+2017,Margutti+2018} at late time $\gtrsim9\,\rm{days}$.
As we pointed out in section \ref{Long-Lasting X-ray Luminosity}, the bound-free opacity is very large and its optical depth is roughly evaluated by \citep{Kisaka+2016} 
\begin{eqnarray}
\tau_{\rm{X}}&\sim&\frac{\kappa_{\rm{X}}M_{\rm{e}}}{4\pi{R_{\rm{in}}^2}}\\
&\sim&2000\,\biggl(\frac{\kappa_{\rm{X}}}{100\,\rm{cm^2\,g^{-1}}}\biggl)\biggl(\frac{M_{\rm{e}}}{0.01\,\Msun}\biggl)\biggl(\frac{v_{\rm{min}}}{0.1\,c}\biggl)^{-2}\biggl(\frac{t}{\rm{day}}\biggl)^{-2},
\end{eqnarray}
where we use the X-ray bound-free opacity of \textit{r-}process elements given in \cite{Hotokezaka+2016}.
Since most X-rays are absorbed until $\sim40\,{\rm{days}}\,(M_{\rm{e}}/0.01\,\Msun)^{1/2}$, the detections by \textit{Chandra} at 9 and 16 days with $L_{\rm{X}}\simeq10^{38-39}\,\rm{erg\,s^{-1}}$ does not constrain the X-ray irradiation luminosity, and it is most likely the afterglow emission.
\cite{2017arXiv171010757M} also discuss high energy emissions from the merger remnant at late time.
They find that if a super Eddington accretion disk with luminosity $L_{\rm{X}}\sim10^{40}\,\rm{erg\,s^{-1}}$ exists as a merger remnant, X-rays from the disk can escape from the ejecta at $30-100\,\rm{days}$ after the merger.
\textit{NuSTAR} gives an interesting upper limit at $\simeq30\,\rm{days}$ \citep{Evans+2017} and rejects the disk luminosity of $L_{\rm{X}}\gtrsim10^{41}\,\rm{erg\,s^{-1}}$.
The late-time X-ray observation \citep{Ruan+2017} also excludes the possibility that the central engine powers the X-ray emission \citep{Margutti+2018}.
On the other hand, the required X-ray irradiation luminosity becomes $L_{\rm{X}}\sim10^{38-39}\,\rm{erg\,s^{-1}}$ at $30-100\,\rm{days}$, which is consistent with these \textit{NuSTAR} and \textit{Chandra} observation. 

In the reprocessed emission model, X-ray and NIR observations may constrain the source geometry through their flux ratio.
In sGRB 130603B, X-ray and NIR excesses had comparable fluxes at 10 days.
If they were powered by the isotropic irradiation from the central engine as we consider here, their comparable fluxes are naturally explained.
Therefore, in this case, the flux ratio does not strongly constrain the source geometry.
For GW170817, if a successful prompt jet evacuated ejecta, no detection of long-lasting X-ray excess suggests an off-axis jet.

\acknowledgments
We thank Masaru Shibata and Masaomi Tanaka for useful comments and discussions.
TM thank Sho Fujibayashi for fruitful discussions in public baths nearby Kyoto University.
TM is partly supported by JSPS Overseas Challenge Program for Young Researchers.
This work is supported by Grant-in-Aid for JSPS Research Fellow 17J09895 (TM) and KAKENHI 24103006, 26247042, 26287051, 17H01126, 17H06131, 17H06357, 17H06362 (KI), and 16J06773 (SK).

\end{document}